# Observation of surface Fermi arcs in altermagnetic Weyl semimetal CrSb

Wenlong Lu[1,†], Shiyu Feng[1,†], Yuzhi Wang[2,3†], Dong Chen[4,5,†], Zihan Lin[1], Xin Liang[1], Siyuan Liu[6], Wanxiang Feng[6], Kohei Yamagami[7], Junwei Liu[8], Claudia Felser[5], Quansheng Wu[2,3]*, Junzhang Ma[1,*]

[1]Department of Physics, City University of Hong Kong, Kowloon, Hong Kong, China
[2]Institute of Physics and Beijing National Laboratory for Condensed Matter Physics, Chinese Academy of Sciences, Beijing 100190, China
[3]University of Chinese Academic of Science, Beijing 101408, China
[4]College of Physics, Qingdao University, Qingdao 266071, China
[5]Max Planck Institute for Chemical Physics of Solids, 01187 Dresden, Germany
[6]Centre for Quantum Physics, Key Laboratory of Advanced Optoelectronic Quantum Architecture and Measurement (MOE), School of Physics, Beijing Institute of Technology, Beijing 100081, China
[7]Japan Synchrotron Radiation Research Institute, 1-1-1, Sayo-cho, Sayo-gun, Hyogo 679–5198, Japan
[8]Department of Physics, The Hong Kong University of Science and Technology, Hong Kong, China.

†The authors contributed equally to this work.
*Corresponding to: Junzhang Ma (junzhama@cityu.edu.hk) Quansheng Wu (quansheng.wu@iphy.ac.cn)

**As a special type of collinear antiferromagnetism (AFM), altermagnetism has garnered significant research interest recently. Altermagnets exhibit broken parity-time symmetry and zero net magnetization in real space, leading to substantial band splitting in momentum space even in the absence of spin-orbit coupling. Meanwhile, parity-time symmetry breaking always induce nontrivial band topology such as Weyl nodes. While Weyl semimetal states and nodal lines have been theoretically proposed in altermagnets, rare reports of experimental observation have been made up to this point. Using ARPES and first-principles calculations, we systematically studied the electronic structure of the room-temperature altermagnet candidate CrSb. At generic locations in momentum space, we clearly observed band spin splitting. Furthermore, we identified discrete surface Fermi arcs on the (100) cleaved side surface close to the Fermi level originating from bulk band topology. Our results imply that CrSb contains interesting nontrivial topological Weyl physics, in addition to being an excellent room temperature altermagnet.**

In condensed matter physics, the lifting of electronic band spin degeneracy in crystalline solids can induce novel physical properties and hold promise for electronic device applications [1,2]. Typically, spin-orbit coupling (SOC) in non-centrosymmetric systems is a key factor driving spin splitting, especially in elements with large atomic numbers [3]. However, this limits the potential material candidates among elements with small atomic numbers. Recent years, theoretical analysis proposed that collinear antiferromagnetic (AFM) order without parity-time symmetry (*PT*) and $\tau U$ symmetry (where $\tau$ is a translation operation and $U$ is a spin rotation operation) can possess dramatic spin splitting without the need to consider SOC [4,5]. This increases the range of material candidates for spin splitting by including compounds made up of small atomic number elements. As a result of their distinct magnetic characteristics, collinear AFM materials have started to garner a lot of attention from researchers [6–17]. Especially interesting is that the intrinsic crystal (*C*-) symmetry ensuring the zero magnetization in AFMs will force contrasting spin polarization at different valley or momenta paired by this crystal symmetry, forming the *C*-paired spin-valley locking [18], which leads to the unique noncollinear spin current generation [18–20] and piezomagnetism [18]. Scientists refer to this kind of AFM arrangement as "altermagnetism" [21–23]. The unique physical characteristics of altermagnets have been verified by lots of recent theoretical and experimental investigations [24–34]. The recent advancements in altermagnets, which include novel physical phenomena associated with altermagnetic candidates, are comprehensively reviewed [35,36]. More discussion about the properties of altermagnetism please see appendix A.

Meanwhile, the breaking of symmetry usually induces novel topological quantum states. Dirac nodes divide into Weyl nodes when parity-time symmetry is broken, giving rise to exotic physical phenomena such as massless quasiparticles, open surface Fermi arcs, chiral anomaly, and the anomalous Hall effect [37–44]. Significant spin splitting is required for an ideal Weyl semimetal in order to generate well-separated Weyl nodes [42–45]. In the majority of perfect topological semimetals, SOC is essential. Scientists have recently realized that altermagnets present a special chance for enormous spin splitting to drive perfect topology [46–48]. Experimental observations on the nontrivial topological band structures of altermagnets are still rare, though, as this topic is still in its beginning. Nontrivial topology combined with altermagnetism may give rise to new physical properties combining the features of altermagnetism and topological physics. Altermagnetic Weyl semimetals may find use in spintronic devices as a result of this synergy.

Calculations have predicted plenty of potential altermagnet candidates [21,22], and recent experimental research has validated several of them [31–33,49,50]. Notably, among these candidates, CrSb shows significant promise, exhibiting large spin splitting up to 1 eV [34]. As previously highlighted, altermagnets hold significant scientific merit and promise for various applications [21,22,25,29]. However, there is still a scarcity of experimental studies on the electronic structures of altermagnets. Electronic bands in CrSb epitaxial thin films were investigated with high photon energy soft X-ray ARPES in recent studies, and the band dispersions indicates an altermagnetic band structure in CrSb [32]. Nevertheless, high resolution band structure, surface states, and topology of CrSb have not yet been thoroughly examined in the low photon-energy region. In this study, using angle-resolved photoemission spectroscopy (ARPES) combined with first-principles calculations, we systematically examined the electronic structure of the room-temperature altermagnet candidate CrSb, employing both soft X-ray and vacuum ultraviolet (VUV) ARPES. We observed clear spin splitting in generic momentum space away from high-symmetry lines. Additionally, we detected distinct surface Fermi arcs on the (100) cleaved side surface near the Fermi level, which emerge from bulk topology. Our results confirm that CrSb is an ideal altermagnetic Weyl semimetal.

*General information of CrSb*. — CrSb hosts a NiAs-type structure with space group $P6_3/mmc$ (#194) [51], as shown in Fig. 1(b), and Fig. 1(c) displays the corresponding 3D Brillouin zone (BZ). Two distinct Sb atoms are interpolated between Cr layers along the c axis and encircled by six Cr atoms. A type AFM order with an in-plane FM structure and an out-of-plane AFM structure with two spin sublattices is formed by the magnetic momentum carried by the Cr atoms. Because of nonequivalent Sb atoms by means of translation operation, the magnetic lattice structure cannot be restored by parity time symmetry. Instead, the two sublattices can be related by the $M_z$ mirror operation, or by the combination of rotation operation $C_{6z}$ and translation operation as shown in Fig.1a. The broken of parity time symmetry and the special sublattice structure define CrSb as g-wave altermagnetism with large spin splitting as demonstrated in Fig.1(a). The X-ray photoemission spectroscopy (XPS) is plotted in Fig. 1(d), with both Cr and Sb elements clearly resolved. The photography of CrSb samples is shown in Fig. 1(e) for reference.

The experimental results of overall bulk band dispersion acquired with soft X-ray along high symmetry lines agree well with the calculations under altermagnetic order in all directions instead of no-magnetic order as shown in Fig.2 (a)-(d). More details please see appendix C. However, soft X-rays have the disadvantage of low energy resolution and statistics and are not sensitive to surface Fermi arcs in the presence of topologically nontrivial states. To address this, we conducted high energy resolution VUV ARPES experiments with photon energies ranging from 30 eV to 148 eV. The photon energy-dependent out-of-plane cut is plotted in Fig. 2(f), which agrees well with the band trend of calculation along the ΓM direction. We determined the high symmetry Γ planes corresponding to a photon energy of 102 eV and M planes at photon energies of 141 eV and 71 eV, which are consistent with the predictions based on the soft X-ray $k_z$ data, as shown in Fig. 2(g).

*Observation of spin splitting*. — After discussing the 3D band structure, we now focus on the band spin splitting, a crucial property for altermagnets. In CrSb, most of the bands along high-symmetry lines are spin double degenerate. To detect the spin splitting, measurements need to be better conducted along generic momentum space. We selected two photon energies, 60 eV and 80 eV, as indicated by the gray planes in the 3D BZ in Fig. 2(e), where the corresponding planes in reciprocal space are both 0.3 times the distance of ΓM away from the M point (Fig. 2(e)). The Fermi surfaces characterized by photon energies of 60 eV and 80 eV are shown in Figs. 3(a) and 3(b), respectively. For

60 eV map, we took three cuts parallel to the ΓK direction from the second BZ, which are respectively evenly 0.075 times the distance of ΓA between each other away from the Γ point: cut1-3 (Fig. 3(d)-(f))). For 80 eV map, as the momentum space covers larger area, we take the cuts with double separation of that from 60 eV map as shown by cut4-5 in Fig. 3(g,h). We found from the experiments that the $k_z$ broadening effect is very large at UV light range which is shown by the black arrows in Fig. 2(f). For each cut, the calculated band structure under magnetism, considering $k_z$ broadening (plot with several sets of data under a range of $k_z$), is shown (Figs. 3(i-m)) for comparison with the ARPES spectra at 60 eV, and 80 eV respectively. In the ARPES intensity results for each cut, clear spin splitting up to 0.2 eV, noted by the corresponding red and blue arrows, is found to be consistent with the spin splitting in the calculated results. This indicates the presence of large spin splitting near the Fermi level under long-range altermagnetism in CrSb, which is also the evidence of parity time symmetry breaking.

*Surface Fermi arcs*. — Besides the bulk bands, we identified bands and Fermi surfaces that do not exist in the bulk band calculations, denoted by green 'SS' in Fig. 3. These bands, located between the projections of Γ and K, form long Fermi surfaces as shown in Figs. 3(a) and 3(b), centered at the surface BZ boundary as shallow hole-like bands indicated by green arrows in Figs. 3(d-h). When checking details with curvature intensity spectra, we identified two bands as shown in the inset of Fig. 3(f). We attribute these bands to surface states, which is further confirmed by the absence of SS bands in the soft X-ray ARPES spectra shown in Fig. 3(c) recorded with 760 eV photon energy.

From the band structure, we can easily identify band inversion and crossings near the Fermi level. Now, our analysis focuses on the topology of CrSb between bands N and N+1 (where N is the occupation number) using the open-source software WannierTools [52]. A preliminary search reveals numerous band crossings, which appear quite complex. In order to simplify these crossing nodes, we categorize them into two sets for spin-up and spin-down bands without SOC, we observed seven nodal rings for each spin subset, indicated by red and blue colors in Figs. 4(a-c). Three nodal rings are located within the gray plane and its equivalents shown in Fig. 4(g) between

$k_z = 0$ and $\pi/c$, with 120 degrees apart from each other and degenerate along the ΓZ high-symmetry line. Another three nodal rings are located between $k_z = 0$ and $-\pi/c$, which can be derived from the first set using the combination of $M_z$ mirror and $C_{6z}$ rotation operations. The seventh nodal ring is located near the $k_z = 0$ plane, exhibiting a wave-like out-of-plane shape. There exists a complex conjugation symmetry, denoted as $C$ (where $C = K$), analogous to time-reversal symmetry in spinless systems. Consequently, the combined antiunitary symmetry of ($C$) and the inversion symmetry ($I$) satisfies (($CI)^2 = 1$), which imposes a constraint that forbids one of the three Pauli matrices in a two-band effective model at any given k-point. This condition ensures the presence of the seventh nodal ring, as described by the topological invariant ($Z_4 = 1$) [53]. The nodal rings for spin-up and spin-down crossing nodes can be related by the $M_z$ mirror operation. When SOC is considered, the nodal rings open gaps except for 12 Weyl points with green spheres representing nodes with chirality $\xi = +1$ and pink spheres representing $\xi = -1$, symmetrically located at $k_z = \pm 0.796\pi/c$ planes, as shown in Figs. 4(d-f). We plot the band structure along a cut that intersects two Weyl points, marked as $-M_W$-$\Gamma_W$-$M_W$ ($k_z = 0.796\pi/c$) in Fig. 4(g). The calculated band structures, both without and with SOC, are displayed in Fig. 4(h), showing the band crossings for nodal rings and Weyl points.

After discussion the topology of bulk bands, we now focus on the surface states. The (100) surface is a natural cleavage plane, as shown in Fig. 4(i). We calculated the band structure projection along this surface, and the calculated surface band structure along high-symmetry lines is shown in Fig. 4(j), with the related Fermi surfaces plotted in Fig. 4(k). Between bands N and N+1, two branches of surface states are clearly resolved along the projection of the $-K'$-$\Gamma'$-$K'$ direction as well as the projection of $-H'$-$A'$-$H'$. These surface Fermi arcs qualitatively agree with our experimental results, as shown in Fig. 3. As the surface potential in reality can be different, the dispersion of surface bands between experiments and calculations always has some differences in detail. Since the projections of Weyl nodes merge in the bulk band projection, it's challenging to determine how the surface Fermi arcs connect with the Weyl nodes. However, from the trend of the surface bands in the middle of Fig. 4(f), it appears that the surface Fermi arcs terminate toward the Weyl nodes. On the (100) cleavage surface, two Weyl nodes with opposite chirality overlap, resulting in two branches of surface Fermi arcs, consistent with the double surface band structure.

In summary, we investigated the complex band structure of the altermagnet CrSb utilizing both UV ARPES and soft X-ray ARPES experiments. We observed a prominent spin-split 3D bulk band structure, proving the predictions of LDA calculations for the g-wave altermagnetism. Furthermore, we have observed surface Fermi arcs on the (100) cleavage that have not been reported in previous studies. Our further analysis by calculations indicates that these Fermi arc states may originate from the bulk topological Weyl nodes which contribute to the unique nontrivial band topology of CrSb. The breaking of parity-time symmetry is crucial in fostering the novel states of altermagnetism and topology within CrSb. This synergy has the potential to enhance the future possible functionality of spintronic devices and expand their diverse applications by combining two sets of unusual physical features of topology and altermagnetism in a single material.

*Note added.—* Recently, we became aware of four preprints regarding the ARPES investigation of CrSb on the (001) top surface [54–57] while preparing the manuscript. Three of them concentrate on the band spin splitting [54–56], while one more talks about the top surface Fermi arcs and topological Weyl semimetal states [57]. There is no published information on the (100) side surface Fermi arc states observed in our study.

The supporting data for this article will be available in a public repository upon publication of the manuscript (MARVEL Materials Cloud Archive) with the same title of this paper (https://archive.materialscloud.org).

This work was supported by National Key R&D Program of China (Grants No.2021YFA1401500), Research Grant Council (RGC) with grant number (21304023, C6033-22G, C1018-22E, 16303821, 16306722 and 16304523), the National Natural Science Foundation of China (12104379, 12274436 and 12274027)), and Guangdong

Basic and Applied Basic Research Foundation (2021B1515130007). C.F. and D.C. thank the European Research Council (ERC Advanced Grant No.742068 `TOPMAT'), the DFG through SFB 1143 (project ID. 247310070) and the Wurzburg-Dresden Cluster of Excellence on Complexity and Topology in Quantum Matter ct.qmat (EXC2147, project ID. 390858490) for support. The SX-ARPES measurement was performed under the approval of BL25SU at SPring-8 (Proposal No.2022A2060, 2022B2106 and 2024A1619). We acknowledge MAX IV Laboratory for time on Beamline Bloch under Proposal 20230361. We acknowledge Dr. Rui Lou for help during the experiments in BESSY, and Dr. Jacek Osiecki, Mats Leandersson, Balasubramanian Thiagarajan for help during experiments in MAXIV, Dr. Kohei Yamagami for help during experiments in Spring8, and Mr. Jianyang Ding and Jiayu Liu for help during experiments in SSRF. We acknowledge Prof. Yuanfeng Xu for helpful discussion.
The authors declare no competing financial interest.

## Appendix A: Discussion about the properties of altermagnetism

Altermagnetism enables unconventional high even-parity wave magnetism, such as d-wave, g-wave, and i-wave, to exist within the framework of an effective single-particle description of magnetism [6,21,22,29]. Recently, it has even been suggested that altermagnetism can extend to include non-collinear spins and multiple local-structure variations with odd-parity spin symmetry [58]. The study of altermagnetism significantly broadens the symmetry classification in the field of magnetism. Altermagnets are promising candidates for driving the next generation of information technology because of their unique physical properties: C-paired spin-valley locking, nontrivial Berry phase, spin currents, the anomalous Hall effect, stability under perturbing magnetic fields, and giant magnetoresistance, etc [4,9,11–16,22,28,59–65].

## Appendix B: Methods

The crystals of CrSb were grown by the self-flux method. Cr and Sb with atomic ratio of 1 : 4 were loaded in an alumina crucible and sealed in an evacuated quartz tube. The tube was heated to 1000 °C, kept for 20 hours, and then cooled down to 750 °C with a rate of 2 °C/h. After that, the sample was taken out of the furnace, and centrifuged to separate the crystals from flux. Conventional ARPES measurements were performed at the beamline UE112 PGM-2b-1^2 of BESSY (Berlin Electron Storage Ring Society for Synchrotron Radiation) synchrotron, at Bloch beam line of MAX-IV synchrotron, and at BL03U beamline of the Shanghai Synchrotron Radiation Facility (SSRF). The energy and angular resolutions were set to ~20 meV and 0.1°, respectively, and the temperature was set to around 20K; Soft X-ray ARPES measurements were performed at BL25SU beamline of SPring-8 Synchrotron and the temperature was set to around 77K, respectively; The samples for all ARPES measurements were cleaved in situ and measured in a vacuum better than $2\times10^{-10}$ Torr. We have calculated the ground state electronic structure of CrSb. Bulk band calculations were performed within the Perdew-Burke-Ernzernhof (PBE) [66] generalized gradient approximation using QUANTUM ESPRESSO package [67,68] with ultra-soft pseudopotential from the PSLibrary [69]. The kinetic energy cutoff for wavefunctions is 80 Ry with a charge density cutoff of 640 Ry. A Monkhorst-Pack [70] $12\times12\times10$ *k*-mesh has been used. For topological band structure calculations in Fig.4, we conducted the ab initio electronic structure calculations for CrSb using the pseudopotential Vienna Ab initio Simulation Package (VASP) [71], with the Perdew-Burke-Ernzerhof (PBE) generalized gradient approximation (GGA) [72]. For the bulk band calculation, a $11 \times 11 \times 9$ k-grid, an energy cutoff of 400 eV, and an energy convergence criterion of $10^{-7}$ eV was employed. For the slab band calculation, a $9 \times 6 \times 1$ k-grid, an energy cutoff of 400 eV, and an energy convergence criterion of $10^{-5}$ eV was employed. To study the topology of CrSb between bands N and N+1, as well as edge states and the Fermi surface along (100), we determined maximally localized Wannier functions using a reduced basis set formed by the d orbitals of Cr, p orbitals of Sb atoms in the Wannier90 software [73] and then used wannhr_symm_Mag [74] to symmetrize the real-space Hamiltonian. The theoretical simulations were conducted using the WannierTools package [52]. The lattice parameters of the primitive cell, a = b = 4.103 Å and c = 5.463 Å.

**Appendix C: General agreement between experimental bulk band structure and calculations**

To confirm the agreement between the bulk experimental band structure and the theoretical calculations, we conducted a systematic soft X-ray ARPES study on CrSb. We cleaved the sample along the (100) side surface. Photon energies versus $k_z$ were easily determined using soft X-ray ARPES data as the $k_z$ resolution is much higher compared with that of low photon energies, which clearly reveals the periodic structure. This periodicity aligns well with the lattice parameter along the normal direction, allowing us to identify the Γ plane at 760 eV and 583 eV, the M plane at 670 eV. The in-plane Fermi surface along the ΓAHK plane is shown in Fig. 2(a), with a rectangular BZ acquired using a photon energy of 760 eV. The photon energy corresponding to the Γ point was confirmed by scanning photon energy and characterizing along the ΓM direction ($k_z$) in momentum space. We present cuts along the high-symmetry lines ΓA, HK, HA, and ΓK in Figs. 2(b) and 2(c), respectively. The calculated band structure under altermagnetic order is plotted alongside for comparison, indicated by red dotted lines. It can be seen that the experimental results agree well with the calculations in all directions. We verified that the materials are under altermagnetic order by plotting the band structure calculations for the non-magnetic order as a reference, indicated by pink dotted lines in Figs. 2(b) and (d), since the magnetic phase transition temperature exceeds ~700 K, which is beyond our experimental setting. The material retains long-range magnetic order under the conditions of our experiment, as indicated by the non-magnetic order calculations not matching the experimental data.

# References

[1] S. A. Wolf, D. D. Awschalom, R. A. Buhrman, J. M. Daughton, S. von Molnár, M. L. Roukes, A. Y. Chtchelkanova, and D. M. Treger, Spintronics: A Spin-Based Electronics Vision for the Future, Science (1979) **294**, 1488 (2001).

[2] I. Žutić, J. Fabian, and S. Das Sarma, Spintronics: Fundamentals and Applications, Rev Mod Phys **76**, 323 (2004).

[3] L. H. THOMAS, The Motion of the Spinning Electron, Nature **117**, 514 (1926).

[4] L.-D. Yuan, Z. Wang, J.-W. Luo, E. I. Rashba, and A. Zunger, Giant Momentum-Dependent Spin Splitting in Centrosymmetric Low-$Z$ Antiferromagnets, Phys Rev B **102**, 014422 (2020).

[5] S. Hayami, Y. Yanagi, and H. Kusunose, Momentum-Dependent Spin Splitting by Collinear Antiferromagnetic Ordering, J Physical Soc Japan **88**, 12 (2019).

[6] Y. Noda, K. Ohno, and S. Nakamura, Momentum-Dependent Band Spin Splitting in Semiconducting $MnO_2$: A Density Functional Calculation, Physical Chemistry Chemical Physics **18**, 13294 (2016).

[7] K.-H. Ahn, A. Hariki, K.-W. Lee, and J. Kuneš, Antiferromagnetism in $RuO_2$ as $d$-Wave Pomeranchuk Instability, Phys Rev B **99**, 184432 (2019).

[8] S. Hayami, Y. Yanagi, and H. Kusunose, Bottom-up Design of Spin-Split and Reshaped Electronic Band Structures in Antiferromagnets without Spin-Orbit Coupling: Procedure on the Basis of Augmented Multipoles, Phys Rev B **102**, 144441 (2020).

[9] L. Šmejkal, R. González-Hernández, T. Jungwirth, and J. Sinova, Crystal Time-Reversal Symmetry Breaking and Spontaneous Hall Effect in Collinear Antiferromagnets, Sci Adv **6**, 23 (2020).

[10] S. A. Egorov and R. A. Evarestov, Colossal Spin Splitting in the Monolayer of the Collinear Antiferromagnet $MnF_2$, J Phys Chem Lett **12**, 2363 (2021).

[11] I. I. Mazin, K. Koepernik, M. D. Johannes, R. González-Hernández, and L. Šmejkal, Prediction of Unconventional Magnetism in Doped $FeSb_2$, Proceedings of the National Academy of Sciences **118**, (2021).

[12] R. González-Hernández, L. Šmejkal, K. Výborný, Y. Yahagi, J. Sinova, T. Jungwirth, and J. Železný, Efficient Electrical Spin Splitter Based on Nonrelativistic Collinear Antiferromagnetism, Phys Rev Lett **126**, 127701 (2021).

[13] M. Naka, Y. Motome, and H. Seo, Perovskite as a Spin Current Generator, Phys Rev B **103**, 125114 (2021).

[14] D.-F. Shao, S.-H. Zhang, M. Li, C.-B. Eom, and E. Y. Tsymbal, Spin-Neutral Currents for Spintronics, Nat Commun **12**, 7061 (2021).

[15] H.-Y. Ma, M. Hu, N. Li, J. Liu, W. Yao, J.-F. Jia, and J. Liu, Multifunctional Antiferromagnetic Materials with Giant Piezomagnetism and Noncollinear Spin Current, Nat Commun **12**, 2846 (2021).

[16] L. Šmejkal, A. H. MacDonald, J. Sinova, S. Nakatsuji, and T. Jungwirth, Anomalous Hall Antiferromagnets, Nat Rev Mater **7**, 482 (2022).

[17] P. Liu, J. Li, J. Han, X. Wan, and Q. Liu, Spin-Group Symmetry in Magnetic Materials with Negligible Spin-Orbit Coupling, Phys Rev X **12**, 021016 (2022).

[18]  H.-Y. Ma, M. Hu, N. Li, J. Liu, W. Yao, J.-F. Jia, and J. Liu, Multifunctional Antiferromagnetic Materials with Giant Piezomagnetism and Noncollinear Spin Current, Nat Commun **12**, 2846 (2021).

[19]  M. Naka, S. Hayami, H. Kusunose, Y. Yanagi, Y. Motome, and H. Seo, Spin Current Generation in Organic Antiferromagnets, Nat Commun **10**, 4305 (2019).

[20]  R. González-Hernández, L. Šmejkal, K. Výborný, Y. Yahagi, J. Sinova, T. Jungwirth, and J. Železný, Efficient Electrical Spin Splitter Based on Nonrelativistic Collinear Antiferromagnetism, Phys Rev Lett **126**, 127701 (2021).

[21]  L. Šmejkal, J. Sinova, and T. Jungwirth, Beyond Conventional Ferromagnetism and Antiferromagnetism: A Phase with Nonrelativistic Spin and Crystal Rotation Symmetry, Phys Rev X **12**, 031042 (2022).

[22]  L. Šmejkal, J. Sinova, and T. Jungwirth, Emerging Research Landscape of Altermagnetism, Phys Rev X **12**, 040501 (2022).

[23]  L. Šmejkal, J. Sinova, and T. Jungwirth, Beyond Conventional Ferromagnetism and Antiferromagnetism: A Phase with Nonrelativistic Spin and Crystal Rotation Symmetry, Phys Rev X **12**, 031042 (2022).

[24]  Z. Feng et al., An Anomalous Hall Effect in Altermagnetic Ruthenium Dioxide, Nat Electron **5**, 735 (2022).

[25]  H. Bai et al., Efficient Spin-to-Charge Conversion via Altermagnetic Spin Splitting Effect in Antiferromagnet RuO2, Phys Rev Lett **130**, 216701 (2023).

[26]  T. Tschirner et al., Saturation of the Anomalous Hall Effect at High Magnetic Fields in Altermagnetic RuO2, APL Mater **11**, 10 (2023).

[27]  Y. Liu, H. Bai, Y. Song, Z. Ji, S. Lou, Z. Zhang, C. Song, and Q. Jin, Inverse Altermagnetic Spin Splitting Effect-Induced Terahertz Emission in RuO2, Adv Opt Mater **11**, 16 (2023).

[28]  L. Šmejkal et al., Chiral Magnons in Altermagnetic RuO2, Phys Rev Lett **131**, 256703 (2023).

[29]  I. I. Mazin, Altermagnetism in MnTe: Origin, Predicted Manifestations, and Routes to Detwinning, Phys Rev B **107**, L100418 (2023).

[30]  B. Brekke, A. Brataas, and A. Sudbø, Two-Dimensional Altermagnets: Superconductivity in a Minimal Microscopic Model, Phys Rev B **108**, 224421 (2023).

[31]  O. Fedchenko et al., Observation of Time-Reversal Symmetry Breaking in the Band Structure of Altermagnetic RuO2, Sci Adv **10**, 5 (2024).

[32]  S. Reimers et al., Direct Observation of Altermagnetic Band Splitting in CrSb Thin Films, Nat Commun **15**, 2116 (2024).

[33]  T. Osumi, S. Souma, T. Aoyama, K. Yamauchi, A. Honma, K. Nakayama, T. Takahashi, K. Ohgushi, and T. Sato, Observation of a Giant Band Splitting in Altermagnetic MnTe, Phys Rev B **109**, 115102 (2024).

[34]  Z. Zhou, X. Cheng, M. Hu, J. Liu, F. Pan, and C. Song, Crystal Design of Altermagnetism, arXiv 2403.07396 (2024).

[35]  L. Bai, W. Feng, S. Liu, L. Šmejkal, Y. Mokrousov, and Y. Yao, Altermagnetism: Exploring New Frontiers in Magnetism and Spintronics, arXiv 2406.02123 (2024).

[36]  M. Hu, X. Cheng, Z. Huang, and J. Liu, Catalogue of C-Paired Spin-Valley Locking in Antiferromagnetic Systems, arXiv 2407.02319 (2024).

[37]  N. P. Armitage, E. J. Mele, and A. Vishwanath, Weyl and Dirac Semimetals in Three-

Dimensional Solids, Rev Mod Phys **90**, 015001 (2018).

[38] X. Wan, A. M. Turner, A. Vishwanath, and S. Y. Savrasov, Topological Semimetal and Fermi-Arc Surface States in the Electronic Structure of Pyrochlore Iridates, Phys Rev B **83**, 205101 (2011).

[39] Z. Wang, Y. Sun, X.-Q. Chen, C. Franchini, G. Xu, H. Weng, X. Dai, and Z. Fang, Dirac Semimetal and Topological Phase Transitions in $A_3$Bi ($A$=Na, K, Rb), Phys Rev B **85**, 195320 (2012).

[40] H. Weng, C. Fang, Z. Fang, B. A. Bernevig, and X. Dai, Weyl Semimetal Phase in Noncentrosymmetric Transition-Metal Monophosphides, Phys Rev X **5**, 011029 (2015).

[41] S.-M. Huang et al., A Weyl Fermion Semimetal with Surface Fermi Arcs in the Transition Metal Monopnictide TaAs Class, Nat Commun **6**, 7373 (2015).

[42] Z. K. Liu et al., Discovery of a Three-Dimensional Topological Dirac Semimetal, Na3Bi, Science (1979) **343**, 864 (2014).

[43] B. Q. Lv et al., Experimental Discovery of Weyl Semimetal TaAs, Phys Rev X **5**, 031013 (2015).

[44] S.-Y. Xu et al., Discovery of a Weyl Fermion Semimetal and Topological Fermi Arcs, Science (1979) **349**, 613 (2015).

[45] D. F. Liu et al., Magnetic Weyl Semimetal Phase in a Kagomé Crystal, Science (1979) **365**, 1282 (2019).

[46] J. Zhan, J. Li, W. Shi, X.-Q. Chen, and Y. Sun, Coexistence of Weyl Semimetal and Weyl Nodal Loop Semimetal Phases in a Collinear Antiferromagnet, Phys Rev B **107**, 224402 (2023).

[47] X. Zhou, W. Feng, R.-W. Zhang, L. Šmejkal, J. Sinova, Y. Mokrousov, and Y. Yao, Crystal Thermal Transport in Altermagnetic RuO2, Phys Rev Lett **132**, 056701 (2024).

[48] D. S. Antonenko, R. M. Fernandes, and J. W. F. Venderbos, Mirror Chern Bands and Weyl Nodal Loops in Altermagnets, arXiv 2402.10201 (2024).

[49] J. Krempaský et al., Altermagnetic Lifting of Kramers Spin Degeneracy, Nature **626**, 517 (2024).

[50] Y.-P. Zhu et al., Observation of Plaid-like Spin Splitting in a Noncoplanar Antiferromagnet, Nature **626**, 523 (2024).

[51] W. J. Takei, D. E. Cox, and G. Shirane, Magnetic Structures in the MnSb-CrSb System, Physical Review **129**, 2008 (1963).

[52] Q. Wu, S. Zhang, H.-F. Song, M. Troyer, and A. A. Soluyanov, WannierTools: An Open-Source Software Package for Novel Topological Materials, Comput Phys Commun **224**, 405 (2018).

[53] Y. Xu, Z. Song, Z. Wang, H. Weng, and X. Dai, Higher-Order Topology of the Axion Insulator EuIn2As2, Phys Rev Lett **122**, 256402 (2019).

[54] G. Yang et al., Three-Dimensional Mapping and Electronic Origin of Large Altermagnetic Splitting near Fermi Level in CrSb, arXiv 2405.12575 (2024).

[55] J. Ding et al., Large Band-Splitting in g-Wave Type Altermagnet CrSb, arXiv 2405.12687 (2024).

[56] M. Zeng et al., Observation of Spin Splitting in Room-Temperature Metallic Antiferromagnet CrSb, arXiv 2405.12679 (2024).

[57] C. Li et al., Topological Weyl Altermagnetism in CrSb, arXiv 2405.14777 (2024).

[58] S.-W. Cheong and F.-T. Huang, Altermagnetism with Non-Collinear Spins, NPJ Quantum Mater **9**, 13 (2024).

[59] M. Naka, S. Hayami, H. Kusunose, Y. Yanagi, Y. Motome, and H. Seo, Spin Current Generation in Organic Antiferromagnets, Nat Commun **10**, 4305 (2019).

[60] R. M. Fernandes, V. S. de Carvalho, T. Birol, and R. G. Pereira, Topological Transition from Nodal to Nodeless Zeeman Splitting in Altermagnets, Phys Rev B **109**, 024404 (2024).

[61] A. Bose et al., Tilted Spin Current Generated by the Collinear Antiferromagnet Ruthenium Dioxide, Nat Electron **5**, 267 (2022).

[62] L. Šmejkal, A. B. Hellenes, R. González-Hernández, J. Sinova, and T. Jungwirth, Giant and Tunneling Magnetoresistance in Unconventional Collinear Antiferromagnets with Nonrelativistic Spin-Momentum Coupling, Phys Rev X **12**, 011028 (2022).

[63] S. Karube, T. Tanaka, D. Sugawara, N. Kadoguchi, M. Kohda, and J. Nitta, Observation of Spin-Splitter Torque in Collinear Antiferromagnetic $RuO_2$, Phys Rev Lett **129**, 137201 (2022).

[64] X. Zhou, W. Feng, X. Yang, G.-Y. Guo, and Y. Yao, Crystal Chirality Magneto-Optical Effects in Collinear Antiferromagnets, Phys Rev B **104**, 024401 (2021).

[65] M. Naka, S. Hayami, H. Kusunose, Y. Yanagi, Y. Motome, and H. Seo, AnAnomalous Hall Effect in $\kappa$-Type Organic Antiferromagnets, Phys Rev B **102**, 075112 (2020).

[66] J. P. Perdew, K. Burke, and M. Ernzerhof, Generalized Gradient Approximation Made Simple, Phys Rev Lett **77**, 3865 (1996).

[67] P. Giannozzi et al., Advanced Capabilities for Materials Modelling with Quantum ESPRESSO, Journal of Physics: Condensed Matter **29**, 465901 (2017).

[68] P. Giannozzi et al., *QUANTUM ESPRESSO:* A Modular and Open-Source Software Project for Quantum Simulations of Materials, Journal of Physics: Condensed Matter **21**, 395502 (2009).

[69] A. Dal Corso, Pseudopotentials Periodic Table: From H to Pu, Comput Mater Sci **95**, 337 (2014).

[70] H. J. Monkhorst and J. D. Pack, Special Points for Brillouin-Zone Integrations, Phys Rev B **13**, 5188 (1976).

[71] G. Kresse and J. Furthmüller, Efficient Iterative Schemes for Ab Initio Total-Energy Calculations Using a Plane-Wave Basis Set, Phys Rev B **54**, 11169 (1996).

[72] J. P. Perdew, K. Burke, and M. Ernzerhof, Generalized Gradient Approximation Made Simple, Phys Rev Lett **77**, 3865 (1996).

[73] A. A. Mostofi, J. R. Yates, Y.-S. Lee, I. Souza, D. Vanderbilt, and N. Marzari, Wannier90: A Tool for Obtaining Maximally-Localised Wannier Functions, Comput Phys Commun **178**, 685 (2008).

[74] Changming Yue, Wannhr_symm_Mag: A Tool for Symmetrization of Magnetic WannierTB, Https://Github.Com/Quanshengwu/Wannier_tools/Tree/Master/Utility/Wannhr_symm_Mag, (n.d.).

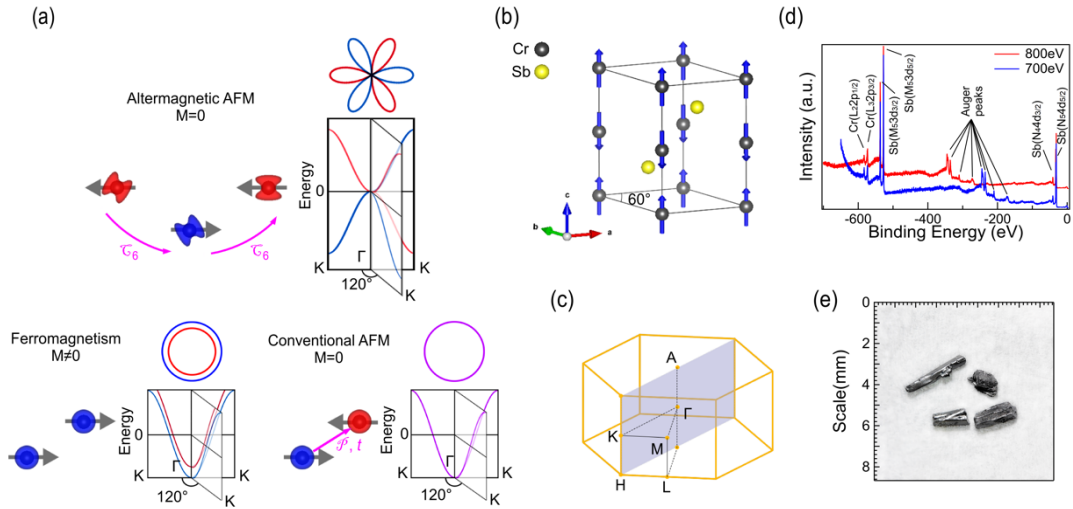

Fig.1. Basic information of CrSb. (a) The schematic illustrating the band spin degeneracy of altermatic AFM, conventional AFM, and FM, respectively. The upper panel represents altermagnetism in hexagonal crystal, such as CrSb, in which the sublattices are correlated by the rotation ($C_{6z}$) operation marked by pink arrows. (b) The crystal structure of the hexagonal CrSb. (c) The corresponding 3D BZ of CrSb, where the shadowed plane runs parallel to the cleavage (100) surface. (d) XPS of CrSb sample, in which each peak of corresponding elements is labeled. (e) The photography of CrSb sample used for experiments.

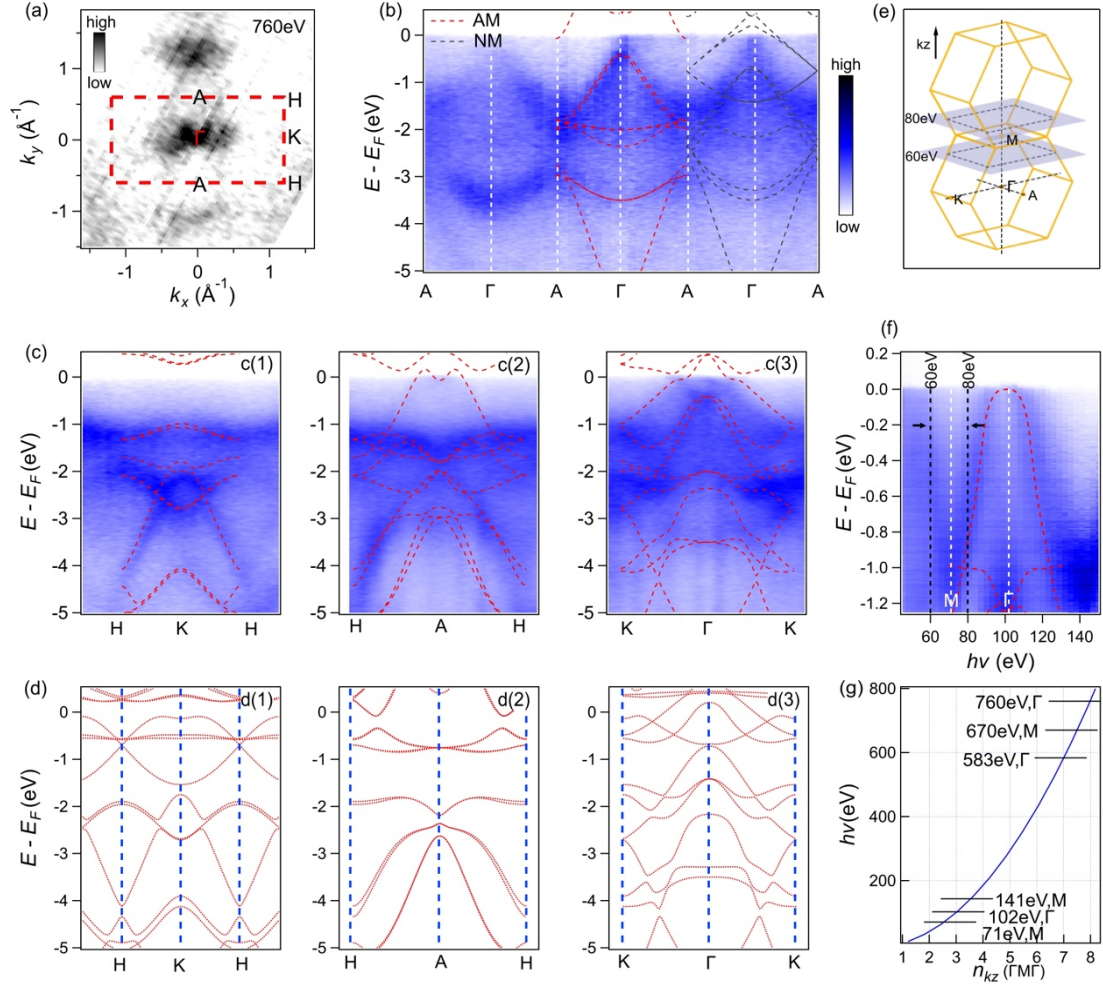

Fig. 2. 3D bulk electronic band structure of CrSb in general. (a) ARPES intensity of the Fermi surface at the K-Γ-A plane acquired with photon energy 760eV. (b) Experimental band dispersion along the Γ-A direction. The calculated band structure with considering SOC under altermagnetic order (red dotted lines) and no-magnetic order (black dotted lines) are plotted alongside for comparison. (c) APRES intensity of cuts along KH, HA, and KΓ high symmetry lines, respectively. The calculated band structure with considering SOC under altermagnetic order (red dotted lines) are plotted alongside for comparison. (d) Calculated band structure under non-magnetic phase with considering SOC along the same high symmetry liens in (c). (e) 3D BZs of CrSb aligned vertically along $k_z$ direction. (f) Photon energy dependent APRES intensity of the cut along the Γ-M direction ($k_z$). The white arrows indicate the substantial $k_z$ broadening effect. (g) The photon energy dependent data fit yielded a relationship curve between photon energy and $k_z$.

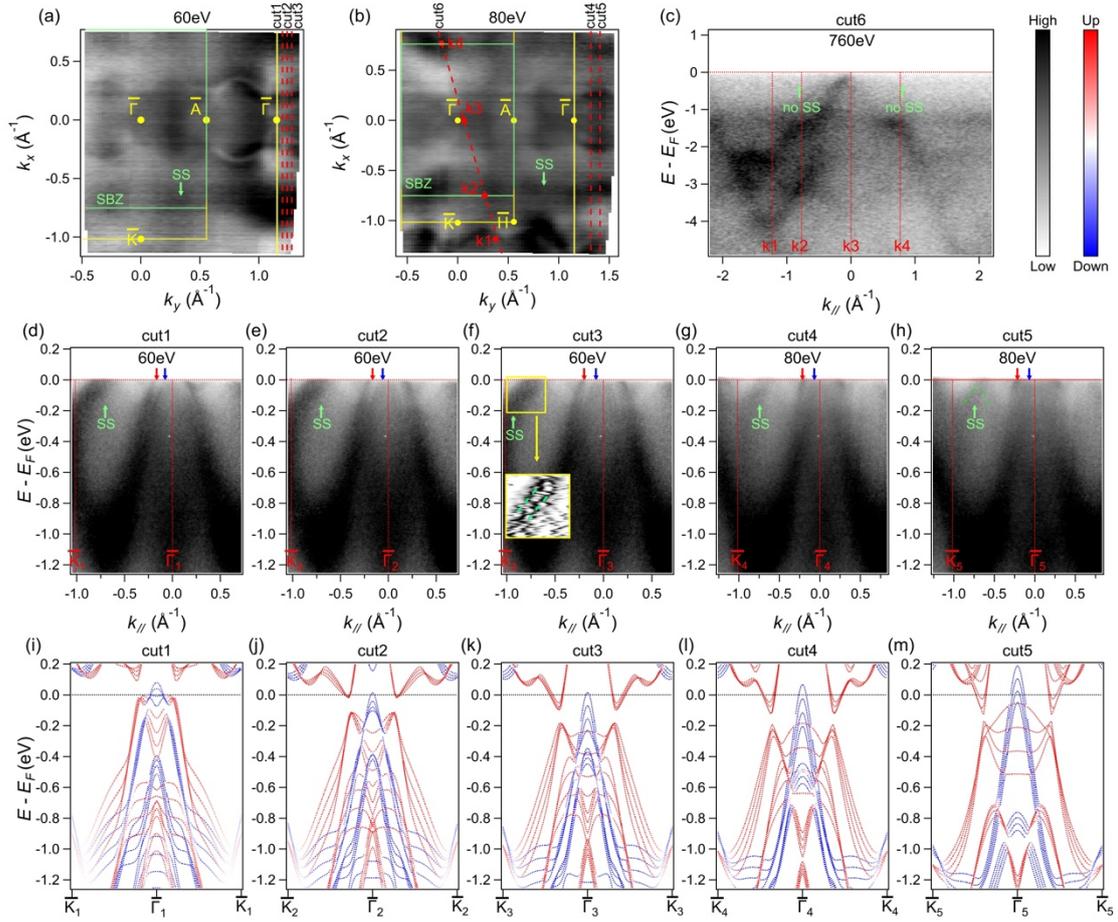

Fig.3. Experimental Fermi surfaces and band splitting of CrSb at arbitrary momentum space. (a,b) ARPES intensity of Fermi surfaces acquired with photon energies 60eV and 80eV, respectively. The yellow line and the green line represent Brillouin zone projection of bulk states and surface states respectively. The red dotted lines indicate the positions of the cut1-5 in the following panels. (c) Soft X-ray data along cut6 without showing any surface bands at the position of the green arrow. (d-h) Experimental band dispersion along cut1-5. The red and blue arrows denote spin splitting. The green arrows denote the surface states. The subpanel in (f) is the curvature intensity plot of the area in yellow box which revealing that the surface states consist of two bands. (i-m) Calculated band structure along cut1-5 with overlapping the results from a range of $k_z$ momentum to simulate $k_z$ broadening in experiments.

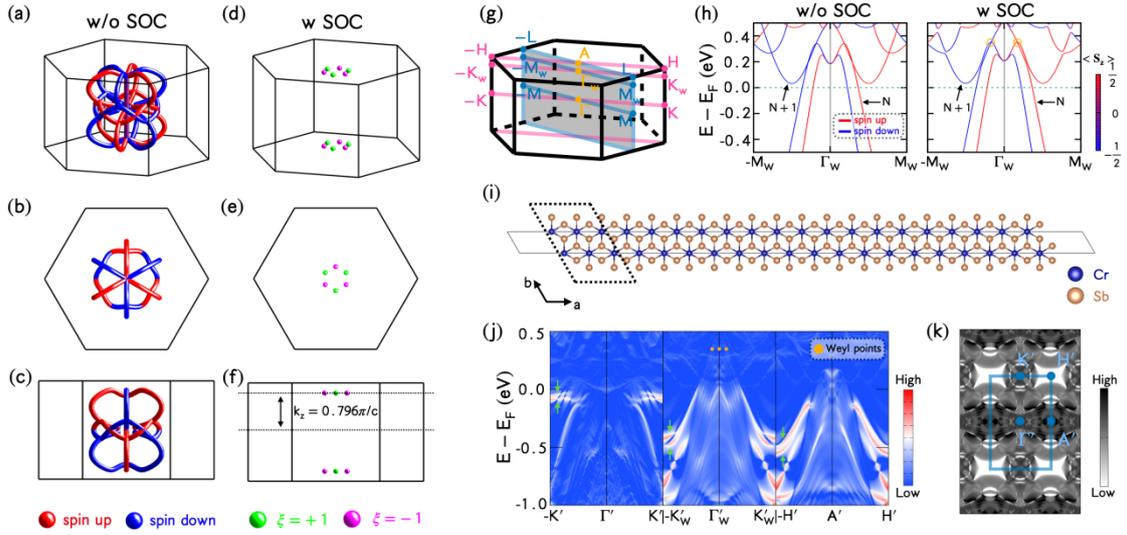

Fig. 4. (a-c) The distribution of the nodal line for bands with occupation numbers N and N+1 (as indicated in (h)) in the first Brillouin zone without SOC, with red (blue) color representing points with spin up (down). (d-f) The distribution of the Weyl points with SOC, with green (pink) spheres representing nodes with chirality $\xi = +1$ (-1). (g) The 3D BZ of the primitive cell of CrSb. (h) Spin splitting bulk bands along the path $-M_W$-$\Gamma_W$-$M_W(k_z = 0.796\pi/c)$ are shown both without and with SOC. The regions highlighted by yellow circles, indicating the positions of the Weyl points. (i) Based on the primitive cell, the structure is expanded by a factor of 20 along the (100) direction with Cr atoms exposed at the surface and the regions enclosed by dashed lines represent the areas projected for our calculations of edge states and the Fermi surface. (j) The calculated (100) surface state along the path $-K'$-$\Gamma'$-$K'|-K'_W$-$\Gamma'_W$-$K'_W|-H'$-$A'$-$H'$. The green arrows point to the surface bands. (k) the Fermi surface of (100), with prime notations representing points in the Brillouin zone of (i).